\documentclass[11pt]{article}
\usepackage[dvips]{graphicx}

\begin{document} 

\title{\Large \bf Nuclear Octupole Correlations and the Enhancement of Atomic 
Time-Reversal Violation} 
\author{ \\ J.\ Engel$^1$, J.L.\ Friar$^2$, and A.C.\ Hayes$^2$ \\ 
{\small $^1$Department of Physics and Astronomy, CB3255,
University of North Carolina,} \\ {\small
Chapel Hill, North Carolina 27599} \\
{\small $^2$Theoretical Division, Los Alamos National Laboratory, Los Alamos, 
New Mexico 87545}}
 
\maketitle

\begin{abstract} 

{\normalsize We examine the time-reversal-violating nuclear ``Schiff moment'' 
that induces
electric dipole moments in atoms.  After presenting a self-contained
derivation of the form of the Schiff operator, we show that the distribution
of Schiff strength, an important ingredient in the ground-state Schiff moment,
is very different from the electric-dipole-strength distribution, with the
Schiff moment receiving no strength from the giant dipole resonance in the
Goldhaber-Teller model.  We then present shell-model calculations in light
nuclei that confirm the negligible role of the dipole resonance and show the
Schiff strength to be strongly correlated with low-lying octupole strength.
Next, we turn to heavy nuclei, examining recent arguments for the strong
enhancement of Schiff moments in octupole-deformed nuclei over that of
$^{199}$Hg, for example.  We concur that there is a significant enhancement
while pointing to effects neglected in previous work (both in the
octupole-deformed nuclides and $^{199}$Hg) that may reduce it somewhat, and
emphasizing the need for microscopic calculations to resolve the issue.
Finally, we show that static octupole deformation is not essential for the
development of collective Schiff moments; nuclei with strong octupole
vibrations have them as well, and some could be exploited by experiment.}

\end{abstract} 

\newpage

\section{Introduction}

\indent

Observation of an atomic electric dipole moment would signal the violation of
time-reversal (T) symmetry\cite{Sachs}, which kaon decay tells us is present
at some level\cite{kaon}.  So far all measurements, whether on elementary
particles or atoms and despite rather high sensitivity, have been
statistically zero, but experiments continue to improve\cite{Lam}.  The level
at which dipole moments are finally seen will help decide among a number of
candidates for the fundamental source of T violation.

Several theorists have proposed that the light actinides would be the best
elements in which to detect a small dipole moment\cite{Haxton,A2,A3}.  Most
recently, the authors of refs.\ \cite{A2,A3} have argued that the existence of
octupole (pear-shaped) deformation in the nuclei of these atoms enhances the
sensitivity of atomic dipole moments to nuclear parity (P) and T violation by
factors of 100 to 1000 (typically about 400 in the later reference) over the
sensitivity in the atom with the best current experimental limit, $^{199}$Hg.
This level of enhancement is due in large part to the existence of close-lying
parity doublets and favorable atomic structure in the light actinides, but
also to fact that it is not the dipole moment of the nucleus that induces a
dipole moment in the surrounding electrons, but rather the ``Schiff moment'',
a quantity that reflects the mean-square radius of the nuclear dipole
distribution.  Asymmetric nuclei have large intrinsic Schiff moments even
though their intrinsic dipole moments are very small, in the same way that a
neutral particle can have a finite charge radius.

The arguments of refs.\ \cite{A2,A3} warrant careful investigation.  In this
paper, we give a pedagogical derivation of the Schiff operator, explore its
action on nuclear ground states, and address the role of octupole correlations
in generating ground-state Schiff moments.  The discussion is organized as
follows:  section I contains a derivation of the nucleus-electron interaction
responsible for atomic-dipole moments and introduces the Schiff operator.  The
section concludes with an evaluation of the nuclear Schiff moment under the
assumption that the dominant source of time-reversal invariance is a nucleon
electric dipole moment.  Section II reveals important differences between the
dipole and Schiff operators, showing that in the Goldhaber-Teller model no
Schiff strength is produced by the giant dipole resonance.  In section III, we
look at Schiff moments in light nuclei, particularly $^{19}$F, confirming the
near absence of Schiff strength in the giant resonance and pointing out a
strong component of strength correlated with low-lying octupole excitations.
The first excited state provides the largest contribution to the Schiff
moment.  Section IV takes up octupole correlations in heavy nuclei, focusing
on octupole deformation.  We find no flaw in the argument that moments in such
nuclei are collective and enhanced, but point to physics that may make the
enhancement less dramatic than claimed in refs.\ \cite{A2,A3} (more detailed
microscopic calculations of both the octupole-deformed nuclei and those that
are currently used in experiments should resolve the uncertainty).  In Section
V we argue that the collective Schiff moments do not depend on the delicate
and sometimes unanswerable question of whether a nucleus is octupole deformed.
Low-lying octupole vibrations generate them in the same way as static octupole
deformation, increasing the number of atoms in which one can expect large
effects.  Section VI summarizes our findings.

\section{Schiff moments}

\indent

We begin by deriving the nucleus-electron interaction responsible for
generating atomic dipole moments.  Though the result is well
known\cite{A3,Khrip,Sushkov}, a complete derivation has never appeared in one
place, and our derivation differs in its details from the others.

In 1939 Feynman\cite{h-f} developed a quantum theory of molecular forces as
part of what is now called the Hellmann-Feynman theorem, or sometimes the
parameter theorem.  This charming relative of the virial theorem\cite{Fock}
allows insight into how forces in a complex quantum system are balanced
against one another, just as they are in a classical system.  In a molecule,
for example, internal Coulomb forces between electrons and nuclei counter each
other so that there is no net force on the molecule (or it would move).  In
the system we consider here, a neutral atom in a uniform electric field, there
is no net charge so there is no net force on the system.  To achieve this the
electrons rearrange themselves \cite{pr} so that there is no net electric 
field at the nucleus (or it would move).

This shielding effect has dramatic and unfortunate implications for
experiments that would probe the atom-nucleus system.  Because nuclei are of
finite extent, however, the shielding of the electrons varies over the nuclear
volume, and this makes probing the nuclear interior possible, as shown
originally by Schiff\cite{Schiff}.  We present next a variant of Schiff's
derivation that uses modern effective-field-theory techniques\cite{power} to
produce the simpler approximate result that has been obtained more
recently\cite{A3,Khrip,Sushkov}.

We assume a neutral (nonrelativistic, for simplicity) atom containing an
extremely heavy nucleus with non-vanishing spin sitting in a uniform electric
field, $\vec{E}_0$.  The atom contains $Z$ electrons, each with charge $e$,
while the nuclear charge is $Z e_p$, where $e_p = -e$ is the proton charge,
and in our units the fine-structure constant is given by $\alpha = e^2_p/4
\pi$.  The nucleus has both an electric monopole distribution and a tiny
electric dipole distribution leading to an electric dipole moment $\vec{d}
\equiv e_p \vec{d}_0$; other moments can be easily added.  In electric-field
gauge the atomic-plus-interacting-nucleus Hamiltonian can be written in the
form
\begin{equation}
\label{eq:atom}
H_{\rm atom} = \sum_{i=1}^Z \left[ K_i + V_i + e \phi (\vec{r}_i) - 
e\vec{E}_0 
\cdot \vec{r}_i \right] -e_p \vec{E}_0 \cdot \vec{d}_0 \ \ , 
\end{equation}
where $K_i$ is the kinetic energy of the i$^{\underline{th}}$ electron
($\vec{p}^{\, 2}_i/2 m_e$ in the nonrelativistic approximation), $\vec{r}_i$ 
and $\vec{p}_i$ are that electron's coordinate and momentum relative to the 
nuclear center-of-mass (CM), $m_e$ is the electron mass,
\begin{equation}
V_i = \alpha \sum_{j < i} \frac{1}{|\vec{r}_i - \vec{r}_j|} 
\end{equation}
is the electron-electron Coulomb interaction, and
\begin{equation}
\phi(\vec{r}_i) = \frac{e_p}{4 \pi} \int \frac{d^3 x \, \rho(\vec{x})}
{|\vec{x} - \vec{r}_i|} 
\end{equation}
is the electrostatic potential due to the complete nuclear charge distribution 
$\rho (\vec{x})$ (with dimensions $\ell^{-3}$, and normalized to $Z$).  

To explore the physics of eq.\ (\ref{eq:atom}) we employ a trick to remove the
last term in that equation.  Because $|\vec{d}_0| \equiv d_0$ is tiny even on
nuclear scales, it is sufficient to manipulate eq.~(1) to first order in that
quantity, ignoring all higher-order terms.  For example, performing a
first-order unitary transformation on $H_{\rm atom}$ produces
$\overline{H}_{\rm atom} \simeq H_{\rm atom} + i[U, H_{\rm atom}],$ where $U
\sim d_0$ in our case is the Hermitian operator
\begin{equation}
U = \frac{\vec{d}_0}{Z} \cdot \sum_{i = 1}^{Z} \vec{p}_i \ \ .
\end{equation}
Performing the commutator generates two terms, one from the last bracketed 
term in eq.~(1) that cancels the interaction of the nucleus and the external 
field ($U$ was constructed to do this), and a second that takes its place:
\begin{equation}
\overline{H}_{\rm atom} = \sum^{Z}_{i=1} \left(K_i + V_i - e \vec{E}_0 \cdot 
\vec{r}_i + e\phi (\vec{r}_i) - \frac{e_p}{Z} \, \vec{d}_0 \cdot 
\vec{\nabla}_i \, \phi (\vec{r}_i) \right) \ \ . 
\end{equation}
The difference, $\Delta$, between eq.~(5) and eq.~(1), which cannot lead to an 
energy shift to first order in $d_0$, is given by
\begin{equation}
\label{eq:delta}
\Delta = e_p \, \vec{d}_0 \cdot \left( \vec{E}_0 - \frac{1}{Z} \sum_{i=1}^Z 
\vec{\nabla}_i \, \phi (\vec{r}_i) \right) = 0 \ \ . 
\end{equation}
This type of relationship, generated by obvious equalities such as $\langle
[U,H_{\rm atom}] \rangle \equiv 0$, is often called a hypervirial
theorem\cite{hyper} and can be derived from the Hellmann-Feynman 
theorem\cite{h-f}.  If
one neglects the finite extent of the nucleus and replaces $\rho(\vec{x})$ 
by $Z \delta^3 (\vec{x})$, eq.~(6) can be rearranged into the form
\begin{equation}
\label{eq:point}
\Delta_{\rm point} = e_p \, \vec{d}_0 \cdot (\vec{E}_0 + \vec{E}_e) \equiv 0 \ 
\ , 
\end{equation}
where $\vec{E}_e$ is the electric field at the nucleus caused by the 
electrons. This simple result states that exact screening holds in the 
point-nucleus approximation, and highlights how a nonzero nuclear volume leads 
to a small but significant breakdown of screening.  Screening is now directly 
incorporated into $\overline{H}_{\rm atom}$.

We can now take advantage of the two very different scales --- atomic and
nuclear --- in the Hamiltonian.  All of the nuclear physics is contained in
$\rho (\vec{x})$ and reflected in $\phi(\vec{r})$; since $(R_N / R_A) \sim
(1\, \mbox{fm}/ 1\, \mbox{\AA}) \sim 10^{-5}$, only a few moments of
$\rho(\vec{x})$ will have practical importance.  For this reason we apply a
derivative expansion of the type used in effective-field theories\cite{power} 
to
$\rho(\vec{x})$.  We assume that the monopole part of $\rho(\vec{x})$ can be
expanded in a series of the form:  $a \, \delta^3(\vec{x}) + b \,
\vec{\nabla}^2 \delta^3 (\vec{x}) + \cdots$.  An analogous expression holds
for the dipole part.  For this to make sense the coefficients $ a, b, \ldots$
together with the derivatives must reflect increasing powers of $R_N/R_A$. 
Moreover, we must preserve conventional definitions, such as
\begin{equation}
\int d^3 x\, \rho(\vec{x})  =  Z \ \ , 
\end{equation}
\begin{equation}
\int d^3 x\, x^2 \, \rho(\vec{x})  =  Z \langle r^2 \rangle_{\rm ch} \ \ , 
\end{equation}
\begin{equation}
\int d^3 x\, \vec{x}\, \rho (\vec{x})  =  \vec{d}_0 \ \ , 
\end{equation}
\begin{equation}
\int d^3 x\, \vec{x}\, x^2 \, \rho (\vec{x})  =  \vec{O}_0 \ \ ,
\end{equation}
where the vector quantity $\vec{O}_0$\cite{ff}, which is the second moment of 
the dipole distribution, bears a similar relationship to $\vec{d}_0$ as $Z 
\langle r^2 \rangle_{\rm ch}$ does to $Z$. Thus, we posit
\begin{eqnarray}
\label{eq:expand}
\rho (\vec{x}) = &\left[ Z \, \delta^3 (\vec{x}) + Z \frac{\langle r^2 
\rangle_{\rm ch}}{6} \vec{\nabla}^2 \, \delta^3 ({\vec{x}}) \right] &- \left[ 
\vec{d}_0 \cdot \vec{\nabla}  \delta^3 ({\vec{x}}) + \frac{\vec{O}_0 \cdot 
\vec{\nabla}}{10} \vec{\nabla}^2 \, \delta^3 ({\vec{x}}) \right] + \cdots 
\nonumber \\ 
\equiv & \rho_{\rm mon} (\vec{x}) &+~~~~~~~~~~~~~~~~~~~
\rho_{\rm dip} (\vec{x})~~~~~~~~~~~~~~+\cdots\ \  
\end{eqnarray}
as a sum of monopole and dipole parts.  Because derivatives with respect to
$\vec{x}$ in $\phi$ (viz., from $\rho$) can be transformed into derivatives 
with
respect to $\vec{r}_i$ through integration by parts, this is indeed an 
expansion
in $R_N/R_A$.

We next separate $\overline{H}_{\rm atom}$ into a part independent of dipole
moments $\vec{d}_0$ and $\vec{O}_0$ and another time-reversal- and 
parity-violating part proportional to these moments: 
\begin{equation}
\overline{H}_{\rm atom} = H^0_{\rm atom} + H_{\rm atom}^{PT} \ \ .
\end{equation}
Expanding $\rho (\vec{x})$ as in eq.\ (\ref{eq:expansion}), we get
\begin{equation}
H^0_{\rm atom} = \sum^{Z}_{i = 1} K_i + V_i - e \vec{E}_0 \cdot \vec{r}_i - 
\frac{Z\alpha}{r_i} + \cdots \ \ , 
\end{equation}
\begin{equation}
H^{PT}_{\rm atom} = - \alpha \sum^{Z}_{i = 1}  \Delta h(\vec{r}_i) \ \ , 
\end{equation}
\begin{equation}
\label{eq:Del}
\Delta h(\vec{r}) = \int \frac{d^3 x \ \rho_{\rm dip} (\vec{x})}
{|{\vec{x}} - {\vec{r}}|} 
 +  \frac{\vec{d}_0 \cdot \vec{\nabla}}{Z}
\int \frac{d^3 x \ \rho_{\rm mon} (x)}{|{\vec{x}} - {\vec{r}}|}  \ \ . 
\end{equation}
Writing out the explicit expansions for $\rho_{\rm mon}$ and 
$\rho_{\rm dip}$ leads to the general result
for $\Delta h$ expressed in terms of the Schiff moment\cite{Sushkov}, 
$\vec{S}$:
\begin{eqnarray}
\label{eq:def}
\Delta h (\vec{r}) &= &4\pi \, \vec{S} \cdot \vec{\nabla} \,
\delta^3 (\vec{r}) + \cdots \nonumber \\ 
\vec{S}& = &\frac{1}{10} \left [ \vec{O}_0 - \frac{5}{3} \vec{d}_0 
\langle r^2 \rangle_{\rm ch} \right ] \ \ . 
\end{eqnarray}
Thus, the coupling of the nuclear dipole distribution to the atomic electrons 
is through the Schiff moment\footnote{The factor of $4 \pi$ in the first of 
eqs.~\ref{eq:def} is often \cite{Lam,Khrip} incorporated in the definition of 
$\vec{S}$.}. The result (\ref{eq:def}) depends in 
leading order on terms of order $R^3_N$, because eq.\ (\ref{eq:point}) 
mandates the cancellation of terms of order $R_N$.

One can make contact with Schiff's paper\cite{Schiff} by defining quantities 
$\rho_C (x)$ and $\rho_M(x)$ such that $\rho_{\rm mon} (x) / Z = \rho_C (x)$ 
and $\rho_{\rm dip} ({\vec{x}}) = -\vec{d}_0 \cdot \vec{\nabla}  \rho_M(x)$. 
Equation (10) of ref.\ \cite{Schiff} then follows from eq.\ (\ref{eq:Del}) 
above: 
\begin{equation}
\Delta h^{\rm Schiff} (\vec{r}) = \frac{\vec{d}_0 \cdot \hat{r}}{r^2} \int d^3 
x \, (\rho_M(x) - \rho_C(x)) \, \theta(r - x) \ \ . 
\end{equation}
This result is exact but not particularly useful. Expanding $\rho_C$ and 
$\rho_M$ in 
the form of eq.\ (\ref{eq:expand}) leads to the approximate result
\begin{equation}
\Delta h^{\rm Schiff} (\vec{r}) \simeq \frac{2\pi}{3} \vec{d}_0 \cdot 
\vec{\nabla} \, \delta^3 (\vec{r}) \left( \langle r^2 \rangle_M - \langle r^2 
\rangle_C \right) \ \ ,  
\end{equation}
from which we deduce his form\footnote{This result is a little misleading 
because it seems to imply that the magnitude of $\vec{O}_0$ should be
$d_0$ times a typical nuclear size; this need not be the case.} of 
$\vec{O}_0$:
\begin{equation}
\vec{O}_0^{\rm Schiff} = \frac{5}{3} \vec{d}_0 \, \langle r^2 \rangle_M \ \ .  
\end{equation}

Finally, the hypervirial (or Hellmann-Feynman) quantity $\Delta $ in eq.~(6) 
can be written using eq.~(12) in the form
\begin{equation}
\label{eq:fh}
\Delta = e_p\, \vec{d}_0 \cdot (\vec{E}_0 + \vec{E}_e + \vec{\nabla}^2 \, 
\vec{E}_e \, \frac{\langle r^2 \rangle_{\rm ch}}{6} + \cdots) \equiv 0 \ \ ,
\end{equation}
where $\vec{E}_e$ is the electrons' electric field at the nuclear CM, and the 
last term arises from averaging that electric field over the nuclear volume.
Because it is the averaged field that cancels the external field at the 
nuclear CM, we see explicitly how nuclear finite size affects screening. 
Moreover, because that last term is equivalent to
\begin{equation}
\label{eq:fhp}
\frac{2 \pi 
\alpha}{3} 
\langle r^2 \rangle_{\rm ch} \sum_{i=1}^Z {\vec{d}}_0 \cdot {\vec{\nabla}}_i 
\, \delta^3 (\vec{r}_i) \ \ , 
\end{equation}
we can see that it produces the $\vec{d}_0$ term in the Schiff moment in eq.\
(\ref{eq:def}) (via the second term in eq.\ (\ref{eq:Del})). The $\vec{O}_0$ 
term in eq.\ (\ref{eq:def}) arises from the first term in eq.\ (\ref{eq:Del}).

Having formulated expressions for nuclear Schiff moments, we want to use them
together with assumptions about the dominant source of P and T violation to
evaluate Schiff moments in real nuclei.  In the rest of this paper we will
assume that a P,T-violating component of the nucleon-nucleon interaction 
causes
a Schiff moment in the distribution of protons, but we conclude this section
by briefly describing another possibility:  that dipole moments of individual 
nucleons are responsible for the nuclear Schiff moment\footnote{Meson-exchange
currents can also generate P,T-violating nuclear moments directly.}.

We introduce proton and neutron (isotopic)
projection operators, $\hat{p}_i$ and $\hat{n}_i$, for the
i$^{\underline{th}}$ nucleon.  The dipole moment must point along the spin,
$\vec{\sigma}_i$, of the nucleon, leading to the impulse-approximation
result\cite{Khrip}
\begin{equation}
\label{eq:nucleon}
\rho_{\rm dip} (\vec{r}) = \left \langle \sum^{A}_{i=1} \hat{p}_i \, d_p \,
\vec{\sigma}_i \cdot \vec{\nabla}_i \, \rho^p_{PT}({\vec{r}}_i - 
{\vec{r}}) + \hat{n}_i \, d_n \, \vec{\sigma}_i \cdot \vec{\nabla}_i \, 
\rho^n_{PT} (\vec{r}_i - \vec{r})  \right \rangle \ \ , 
\end{equation}
where $\rho^p_{PT}$ and $\rho^n_{PT}$ are the proton and neutron 
electric dipole densities (normalized to 1) associated with $d_p$ and $d_n$.  
This yields
\begin{equation}
\vec{d}_0 = \left \langle \sum^{A}_{i=1} \left( \hat{p}_i \, d_p + \hat{n}_i 
\, d_n \right) \vec{\sigma}_i \right \rangle \equiv \vec{d}^{\,p}_0 + 
\vec{d}^{\,n}_0 \ \ , 
\end{equation}
which depends only on the ground-state expectation value of nucleon spin and 
isospin operators.  The Schiff moment can be obtained by evaluating 
$\vec{O}_0$  with eq.\ (\ref{eq:nucleon}), producing
\begin{equation}
\frac{\vec{O}_0}{10} = \frac{\vec{d}^{\,p}_0}{6} \left( \langle r^2 
\rangle^p_{PT} + \langle r^2 \rangle^Z_{PT} \right) + 
\frac{\vec{d}_0^{\,n}}{6} \left (\langle r^2 \rangle_{PT}^n + \langle r^2 
\rangle^N_{PT} \right) \ \ , 
\end{equation}
where $\langle r^2 \rangle^p_{PT}$ and $\langle r^2 \rangle^n_{PT}$ 
are the mean-square radii of the densities $\rho^p_{PT}$ and $\rho^n_{\rm 
PT}$ and
\begin{equation}
\left \langle \sum_{i=1}^A \hat{p}_i \, d_p \, \vec{\sigma}_i \, r_i^2 
\right \rangle \equiv \vec{d}_0^{\,p} \langle r^2 \rangle_{PT}^Z \ \ ; \ \ 
\ \ \ \ \ \left \langle \sum_{i=1}^A \hat{n}_i \, d_n \, \vec{\sigma}_i \, 
r_i^2 \right \rangle \equiv \vec{d}_0^{\,n} \langle r^2 \rangle_{PT}^N \ \ 
. 
\end{equation}
We expect $\langle r^2 \rangle_{PT}^{Z,N}$ to be comparable to nuclear sizes, 
and thus much larger than $\langle r^2 \rangle_{PT}^{p,n}$, which should be 
comparable to nucleon sizes.

\section{Distribution of Schiff strength: its significance and 
the role of the giant dipole resonance} 

\indent

For our considerations, as we have said, the most important mechanism for a 
nuclear ground-state electric dipole or Schiff moment is the action of a 
pseudoscalar T-violating nucleon-nucleon potential, $\hat{V}_{PT}$. This 
induces a Schiff moment given by second-order perturbation theory:
\begin{equation}
\label{eq:SM}
S \equiv \langle S_z \rangle = \sum_{i \neq 0} \frac {\langle \Psi_0 
|S_z|\Psi_i \rangle \langle 
\Psi_i | \hat{V}_{PT} | \Psi_0 \rangle} {E_0 - E_i} + c.c. \ \ , 
\end{equation}
where the state $| \Psi_0 \rangle$ has $J_z = J \neq 0$ and $S_z$ is the
$z$-component of the Schiff-moment operator $\vec{S}$.  The distribution of
Schiff strength to the excited states $| \Psi_i \rangle$ is therefore a
crucial ingredient in the ground-state moment.  At first sight, one might
think that this distribution should resemble that of the electric dipole
operator $\vec{r}\tau_z$, but the two are remarkably different.  Most of the
electric dipole strength is in a broad resonance at 10 to 20 MeV of excitation
energy.  Almost none of the Schiff strength, however, goes to the states in
the giant resonance.  In fact, in the simple but venerable Goldhaber-Teller
(GT) model, in which the giant dipole resonance corresponds to the $1 \hbar
\omega$ oscillation of all the protons with respect to all the neutrons, the
Schiff strength to the resonance is identically zero.  

To see this, we start by assuming\cite{GT} rigid distributions for the $Z$
protons and for the $N$ neutrons, which oscillate harmonically about their
CM with frequency $\omega$.  The separation of the two rigid
spherical (for simplicity) distributions is denoted $\vec{q}$ and the distance
of the proton CM from the overall CM is $N\vec{q}/A$.  The charge distribution
is then given by
\begin{equation}
\label{eq:rhogt}
\hat{\rho}_{\rm GT} (\vec{x}) = \rho_0 \left (| \vec{x} - N\vec{q}/A| \right) 
\ \ ,
\end{equation}
where $\rho_0$ is the ``bare'' nuclear-ground-state charge distribution
normalized to $Z$ protons.  Expanding this equation, one finds components of
$\hat{\rho}_{\rm GT}$ proportional to $|\vec{q}|$, $|\vec{q}|^2$,
$|\vec{q}|^3$, $\cdots$, or equivalently (because $\vec{q}$ oscillates
harmonically), $1 \hbar \omega$, $0$ and $2 \hbar \omega$, $1$ and $3 \hbar 
\omega$, etc.  In particular one finds that the $\vec{q}^{\,2}$-term 
renormalizes the ground-state charge distribution through ``vacuum 
fluctuations'' (viz., the {\em net} 0$\hbar \omega$ part, where the nucleus is 
excited 1 $\hbar \omega$ and then deexcited by the same amount), a result that 
we will neglect for the moment since it doesn't affect the basic physics.
Ignoring all but the monopole and dipole parts in eq.\ (\ref{eq:rhogt}), we 
have
\begin{equation}
\label{eq:rhogtexp}
\hat{\rho}_{\rm GT} \simeq \rho_{\rm 0} (x) - \frac{N}{A} {\vec{q}} \cdot 
\vec{\nabla} \rho_{\rm 0} (x) - \frac{N^3}{10 A^3} \ \vec{q} \cdot 
\vec{\nabla}
\, \vec{q}^{\,2} \, \nabla^2 \rho_{\rm 0} (x)|_{3 \hbar \omega} + \cdots \ \ .
\end{equation}
We perform the usual decomposition of $\vec{q}$ in terms of normalized
(Cartesian) creation and destruction operators
\begin{equation}
\vec{q} = (\vec{a}^{\dagger} + \vec{a}) \sqrt{\frac{\hbar}{2\mu \omega}} ~,  
\end{equation}
with
\begin{equation}
[a_i, a^{\dagger}_j] = \delta_{i j} \ \ ,  
\end{equation}
where $\mu^{-1} = [Z m]^{-1} + [N m]^{-1}$ is the inverse reduced mass of the 
(rigid) protons-neutrons system, and $m$ is the nucleon mass. 

We discuss the effects of vacuum fluctuations under ref.\ \cite{bch} in the
reference list.  The only modification they produce is the replacement of
$\rho_0$ in eq.\ (\ref{eq:rhogtexp}) by the complete ground-state charge
density of the model, $\rho_{\rm ch}$.  This ``renormalization'' (removal of 
the
vacuum fluctuations) has for example shifted the 1$\hbar \omega$ component of
the operator $q^2\, \vec{q}$ (the last term in eq.~(\ref{eq:rhogtexp})) to the 
second term of that equation, and only true 3$\hbar \omega$ excitations remain
from that operator. The modified eq.~(\ref{eq:rhogtexp}) then expresses the
nuclear charge-density operator in terms of the ground-state charge density, 
the 1$\hbar \omega$ transition charge density, the 3$\hbar \omega$ (dipole)
transition density, etc.

The contribution of the 1$\hbar \omega$ excitations to the nuclear moments
$\vec{d}_0$, and $\vec{O}_0$ are now easy to obtain. To evaluate the matrix 
element of the Schiff operator in eq.\ (\ref{eq:SM}) we need these two moments 
of the transition charge density, defined by the second term in eq.\ 
(\ref{eq:rhogtexp}) ($\rho_{\rm dip} (\vec{x}) = - [N/A]\, \vec{q} \cdot
\vec{\nabla} \rho_{\rm ch} (x)$ when vacuum fluctuations are included), 
expressed in terms of $\vec{q}$, which contains the nuclear raising and 
lowering operators. One determines the 1$\hbar \omega$ components of these 
moments from eqs.~(8-11):
\begin{equation}
\label{eq:d}
\vec{d}_0^{\: \rm GT} = \frac{Z N}{A} \vec{q} \ \ , 
\end{equation}
\begin{equation}
\label{eq:O}
\vec{O}_0^{\: \rm GT} (1 \hbar \omega) = \frac{5}{3}\, \vec{d}_0^{\: \rm GT}\,
\langle r^2 \rangle_{\rm ch} \ \ . 
\end{equation}
The electron-nucleus coupling is defined by $\Delta h (\vec{r})$ in eq.\ 
(\ref{eq:Del}). Recalling that $\rho_{\rm mon}$ in that equation is just 
$\rho_{\rm ch}$ here, and using eq.\ (\ref{eq:d}) in $\rho_{\rm dip}$, we 
immediately obtain
\begin{equation}
\label{eq:h=0}
\Delta h_{\rm GT} (1 \hbar \omega) \equiv 0 \ \ ,
\end{equation}
implying that 
\begin{equation}
\label{eq:S=0}
\vec{S}_{\rm GT} (1 \hbar \omega) \equiv 0 \ \ . 
\end{equation}
The last conclusion also follows directly from eqs.\ (\ref{eq:O}) and 
(\ref{eq:def}).  Equation (\ref{eq:h=0}) is more general than eq.\ 
(\ref{eq:S=0}), however,  because it is true to all orders in $R_N/R_A$. 

Thus, despite the fact that it contains all the dipole strength, the
Goldhaber-Teller giant resonance generates no contribution to the Schiff
moment and therefore to the atomic dipole moment.  In the next section we will 
see that the same is nearly true in real nuclei.

\section{Schiff-strength distributions, octupole correlations, and Schiff 
moments in light nuclei.}

\indent

To understand in greater detail the distribution of Schiff strength and the
resulting ground-state Schiff moment, we first examine the situation in light
nuclei.  Although the small radii and charges of light nuclei mean that their
Schiff moments will not be large compared to those of heavy nuclei, they have
the advantage that their structure can be calculated at a detailed microscopic
level.

What kinds of excitations will carry the Schiff strength?  We describe
elementary excitations in terms of harmonic-oscillator shell-model quanta 
($\hbar \omega$); these are not exactly the same as the $\hbar \omega$ of the
Goldhaber-Teller model, but the two are related.  As shown by
Brink\cite{brink}, the electric dipole operator $\vec{d}_0$ can excite only
those components of the harmonic-oscillator shell-model Hamiltonian (with no
residual interactions) corresponding to the giant resonance.  That is, the
simple 1$\hbar \omega$ electric dipole excitations in the harmonic oscillator
are exactly the same as in the Goldhaber-Teller model.  The operator
$\vec{O}_0$, on the other hand, can excite other shell-model modes ---
isoscalar 1$\hbar \omega$ and all kinds of 3$\hbar \omega$ --- that are not a
part of the GT model.  The GT part of the 1$\hbar \omega$ excitations will
cancel in $\vec{S}$ as shown above; the isoscalar 1$\hbar \omega$ and all
3$\hbar \omega$ excitations, by contrast, will contribute to $\vec{S}$.

Octupole modes are important because the E3 and $\vec{O}_0$ operators are in
some sense the $L=3$ and $L=1$ angular-momentum projections of the same
operator.  The {\it isoscalar} $O_0$ and E3 strengths, which contain both
1$\hbar \omega$ and 3$\hbar \omega$ components, are pulled down into low-lying
$1^-$ and $3^-$ states (in even-even nuclei) with similar structure.  In
combination with the suppression of the Schiff strength in the giant-resonance
region, this similarity in structure results in most of the available Schiff
strength being strongly correlated with octupole excitations.  The correlation
can be seen in the closed-shell nucleus $^{16}$O even without much
calculation.  The lowest 3$^-$(6.05 MeV; T=0) state in $^{16}$O has an
enhanced E3 transition to the ground state, B(E3)=13.5$\pm$0.7 Weisskopf units
(W.u.).  The lowest 1$^-$(7.12 MeV;T=0) state decays to this 3$^-$ state
through an enhanced E2 transition, B(E2)=$21\pm5$ W.u., suggesting that the
two states are of similar structure (i.e.\ that the $1^-$ state is a
quadrupole phonon coupled to the 3$^-$ state).  The $1^-$ state shows an
enhanced isospin-forbidden E1 transition to the ground state,
B(E1)=$(3.6\pm0.4)\times 10^{-4}$ W.u.  Although isospin mixing obviously
contributes to the transition, a significant portion of the isospin-forbidden
E1 strength appears to come from a large isoscalar matrix element of
$\vec{O}_0$, a part of the E1 operator that is normally masked\cite{dan}.

Shell-model calculations reflect the strong correlation between the lowest
isoscalar $3^-$ and $1^-$ states \cite{Agassi,Millener}.  Furthermore, the 
$2p-2h$ ground-state correlations in such calculations have a large overlap
with the state formed by acting on the closed shell with two successive E3
operators.  These higher-$\hbar\omega$ correlations in the ground-state wave
function enhance the excitation of octupole-like $3\hbar\omega$ components in 
the 1$^-$ and 3$^-$ states, leading to larger low-lying isoscalar E3 and 
$\vec{O}_0$ matrix elements.

Let us see how this physics works out in $^{19}$F, which has odd A and is
therefore able to have a ground-state Schiff moment.  The ground state of
$^{19}$F is the $1/2^+$ head of a $K = 1/2^+ (sd)^3$ rotational band, while
the first excited state (a 1/2$^-$ state at 110 keV) is the basis for a 4p-1h
($p^{-1}(sd)^4$) $K=1/2^-$ band.  SU(3)-basis shell-model 
calculations\cite{harvey}
for these two bands indicate that the $1/2^-$ band is an octupole excitation
of the ground-state band (in fact $^{19}$F is considered the closest thing
among light nuclei\cite{Butler} to an octupole-deformed system, though a
nonnegligible part of the $1/2^-$ state is an isovector excitation).  Though
this relation in itself implies large E3 and Schiff matrix elements between
these two bands, we have to go beyond these old restricted calculations to get
the full picture, for two reasons.  First, the just-mentioned octupole 
ground-state correlations that further enhance the Schiff strength were
omitted.  Second, we need to ensure realistic behavior for the strength
distribution of the P,T-violating NN interaction, which according to eq.\
(\ref{eq:SM}) is as important as the Schiff-strength distribution in
determining the ground-state Schiff moment.  Fortunately, its behavior is much
simpler.  For the sake of pedagogy, we consider in this section a simple but
often accurate\cite{town} one-body approximation to $\hat{V}_{PT}$:  ${\rm
const} \times \vec{\sigma} \cdot \vec{r}$ in both isovector and isoscalar
channels (we will use a more sophisticated one-body approximation later).  The
$1\hbar\omega$ E1 and $\hat{V}_{PT}$ operators then differ only in their
effect on spin (they are in the same SU(4) multiplet).  The strength from the
isovector $\vec{\sigma} \cdot \vec{r}$ is therefore concentrated in an SU(4)
analog of the giant dipole resonance, and to first approximation lies at the
same energy.  The isoscalar strength is also mostly in a resonance at a
similar energy, so that low-lying P,T-violating strength is
depleted\cite{vogel}.  There will be some strength at low energies where the
Schiff strength is concentrated, just as there is some E1 strength, but it
will represent the tail of a resonance.  This tail was not included in the
calculations of ref.\ \cite{harvey}.

\begin{figure}[htb]
\begin{center}
\includegraphics[angle=-90,width=14cm]{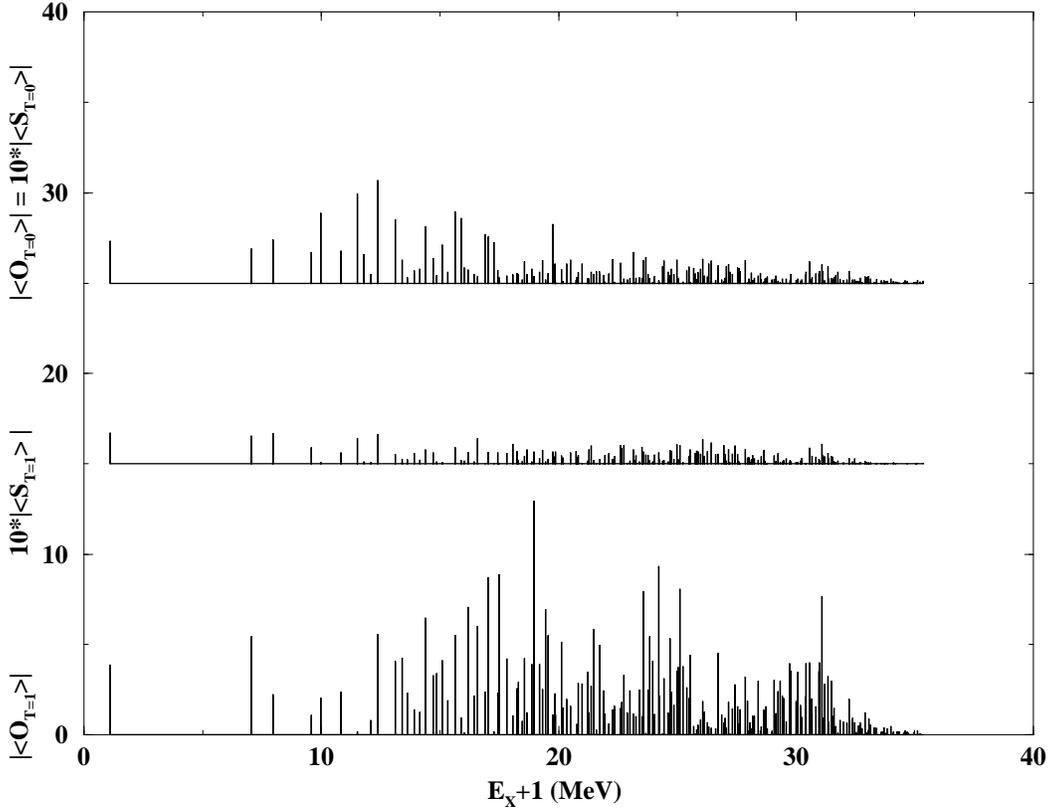}
\end{center}
\caption[Figure 1]{The distribution of isovector ${O}_0$ strength 
(bottom), isovector Schiff strength (middle) and isoscalar Schiff ($\propto 
O_0$) strength (top) in a shell-model calculation of $^{19}$F.  See text for 
discussion.}
\label{f:1}
\end{figure}

Figures 1-3 display the results of a complete (0+1)$\hbar\omega$ shell-model
calculation for $^{19}$F, with the center-of-mass motion fully eliminated.  In
these calculations contributions to the Schiff moment from 2 and
3$\hbar\omega$ excitations have been included through the use of an effective
charge\footnote{It is difficult to treat the important $2\hbar\omega$ and
$3\hbar\omega$ excitations consistently in the low-lying states.  However, the
concept of an effective charge for E3 transitions works well throughout this
mass region.  The results of very truncated (0+1+2+3)$\hbar\omega$
calculations suggest a similar prescription for the $\vec{O}_0$ operator, and
in our $1\hbar\omega$ calculations we applied octupole effective charges to
$\vec{O}_0$.  Reference \cite{dan} makes a case for this same kind of
renormalization.}.  Figure 1 shows the isovector ${O}_0$ distribution, and
above it the isovector $S$ strength.  The extent of the cancellation suggested
by the Goldhaber-Teller model is remarkable.  Also displayed is the isoscalar
Schiff strength, which is uncancelled and has a significant low-lying
component (again correlated with the E3 distribution).  Figure 2 displays the
calculated $\vec{\sigma} \cdot \vec{r}$ strength, and the giant resonances
(including tails) are evident in both the isovector and isoscalar channels.
Finally, in fig.~3 we graph the terms in eq.\ (\ref{eq:SM}) as a function of
excitation energy, assuming that the isoscalar and isovector potentials have
equal strength; from this it is clear that the lowest $1/2^-$ state almost
completely determines the Schiff moment, which is given by the sum of all the
lines in the plot.

\begin{figure}[hbt]
\begin{center}
\includegraphics[angle=-90,width=14cm]{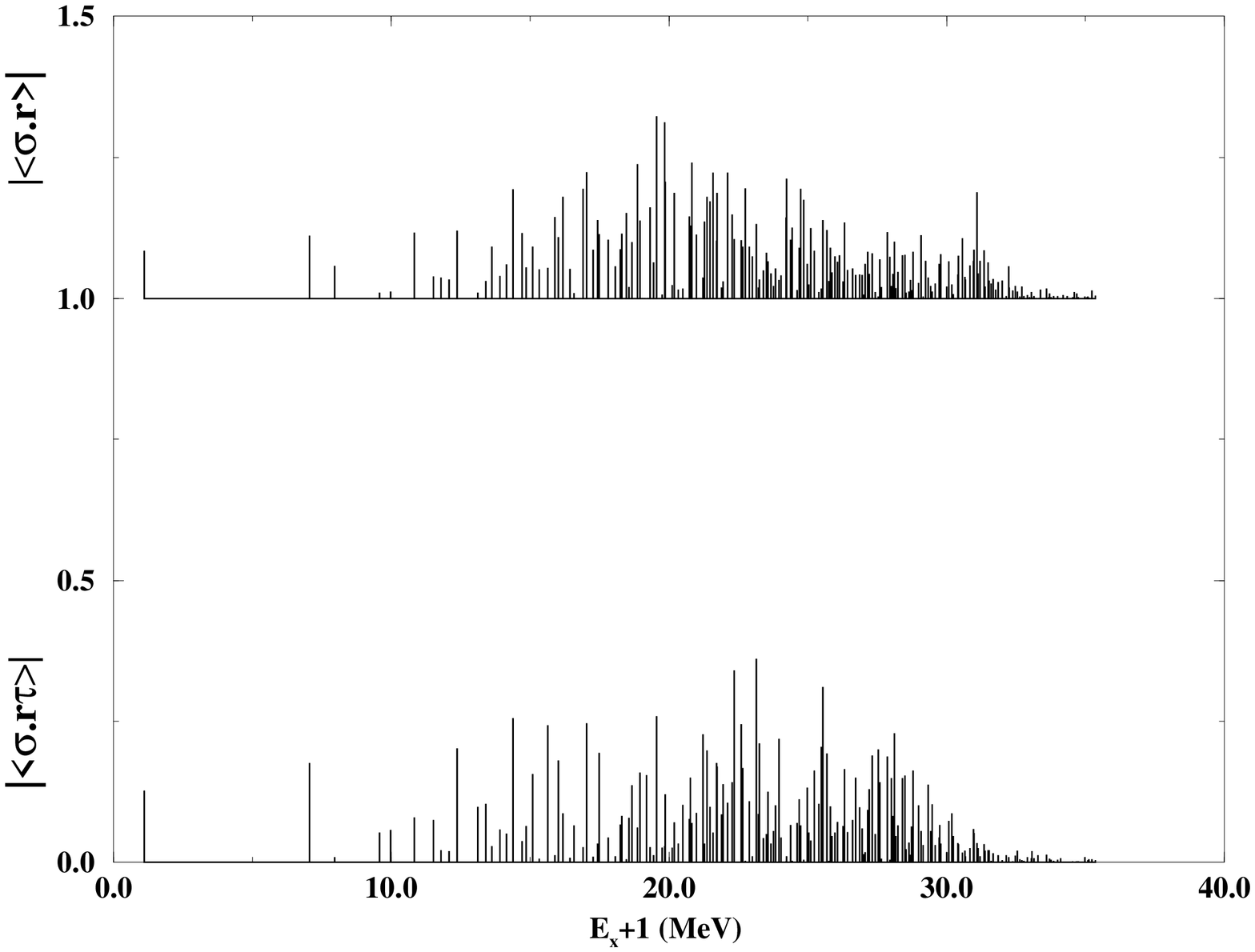}
\end{center}
\caption[Figure 2]{Distributions of the isovector (bottom) and isoscalar (top) 
$\vec{\sigma} \cdot \vec{r}$ strengths in $^{19}$F.  The corresponding 
operators are approximations to the P,T-violating nuclear potential.}
\label{f:2}
\end{figure}
\begin{figure}[tb]
\begin{center}
\includegraphics[angle=-90,width=14cm]{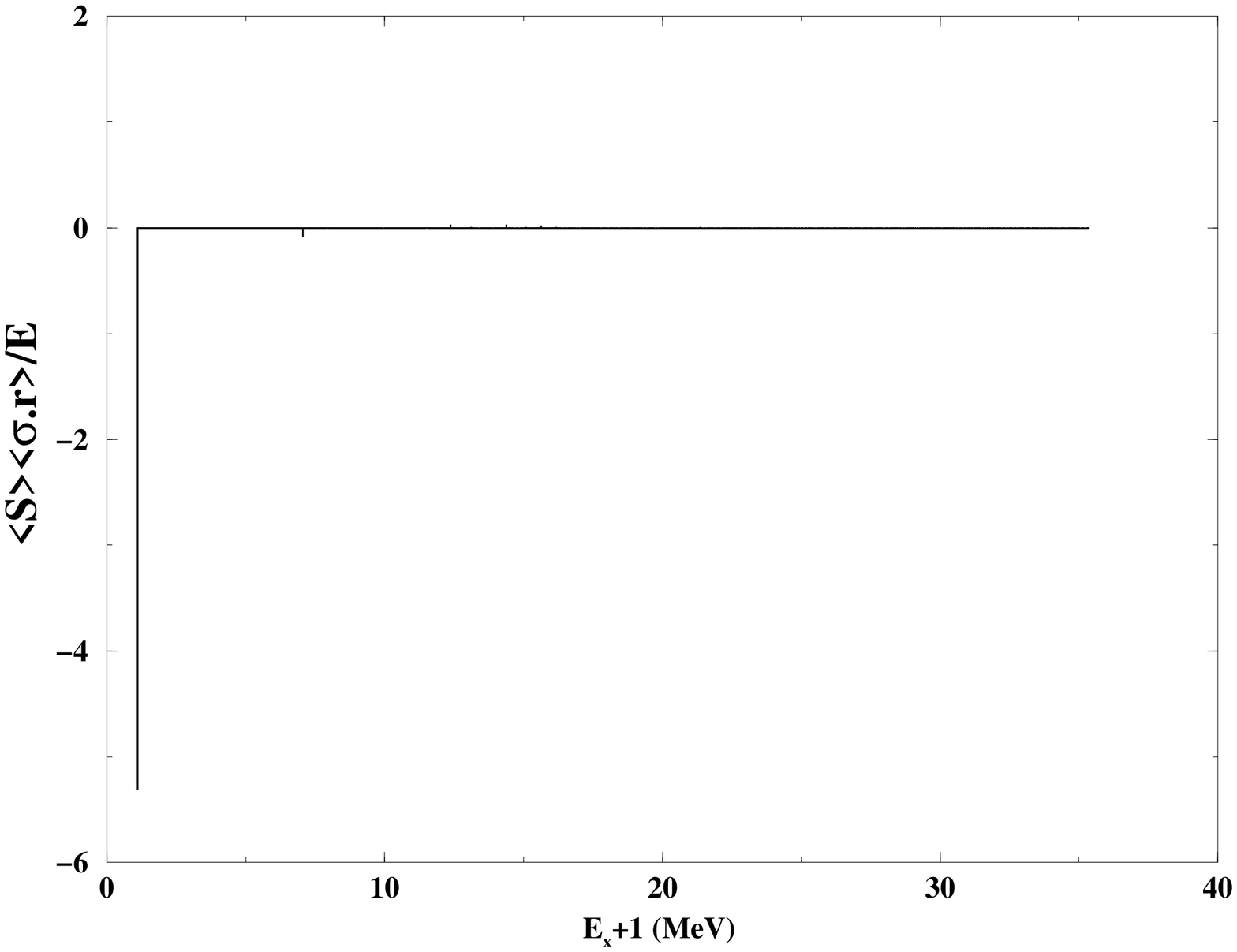}
\end{center}
\caption[Figure 3]{The contributions of individual states in $^{19}$F to the 
ground-state Schiff moment through eq.\ (\ref{eq:SM}).  The first excited 
state is dominant.}
\label{f:3}
\end{figure}

We expect the gross features of the Schiff and $V_{PT}$ distributions to be
general.  The Schiff strength will be correlated with the E3 strength and lie
low in energy.  Nuclei with octupole deformation, where the E3
strength lies as low as $\sim$50 keV and is particularly concentrated, will
show the most-enhanced Schiff moments.  Nuclei with strong low-lying octupole
vibrations should also show enhancement.  As the E3 strength moves
up in energy (and/or the octupole collectivity is diluted), so should the 
dominant contribution to the Schiff strength, and
the Schiff moments will become smaller.  In the remainder of this paper we
examine the extent to which these statements are true in heavy nuclei, where
shell-model calculations are not possible.

\section{Collective Schiff moments in heavy octupole-deformed nuclei.}

\indent

We begin our discussion with a short review of the arguments of ref.\
\cite{A3}, the crux of which is related to what we have already discussed: the
ground-state Schiff moment need not be directly related to the dipole moment,
and in the case of octupole-deformed nuclei is considerably more enhanced.
The authors adopted the particle-rotor model, which is not fully microscopic
and omits a certain amount of valence-space physics.  The arguments are based
on collective octupole correlations, however, and this model represents them
clearly and efficiently.

In the particle-rotor model the nucleus is described as a single particle
coupled to a collective core, the shape of which can be specified through a
function that describes the dependence of its radius on angle:
\begin{equation}
R(\theta,\phi) = R_0 \left(1 + \sum_{l,m} (-1)^m \alpha_{l,m} Y_{l,-m} 
\right)~,
\end{equation}
where $R_0 = 1.2 A^{1/3}$.  In the intrinsic frame of a deformed
nucleus three of the
$\alpha$ variables are no longer independent; they are replaced by three Euler
angles $\theta_i$ ($i=1,2,3$) that specify the orientation with respect to the
laboratory.  We will assume axial symmetry so that all intrinsic $\alpha$'s
vanish except for the $\alpha_{l,0}$'s, which we denote by $\beta_l$.  The
valence particle, in the ``strong-coupling'' version of the model, moves in a
potential that is deformed to match the shape of the core.  The full nuclear
wave function (for a state with angular momentum quantum numbers $J$ and $M$, 
intrinsic magnetic quantum number $K$, 
and parity $p$) thus depends on the Euler angles, the intrinsic deformation 
parameters $\beta_l$, and the intrinsic space ($\vec{r}$) and spin ($s$) 
coordinates of the odd particle:
\begin{equation}
\label{eq:wf}
\langle \theta_i,\beta_l,\vec{r},s |\Psi_{JMK,p} \rangle = 
{\cal N} \left[ 1+\hat{R}_2\right] D^{J*}_{MK}(\theta_i) 
\left[1+p\hat{P}\right] \langle \beta_l,\vec{r},s |\Psi_{\rm int} \rangle ~,
\end{equation}
where the intrinsic particle-core state factorizes as\footnote{In axially 
symmetric nuclei, the component $K$ of total angular momentum  along the 
symmetry axis is equal to that of the particle's angular momentum (sometimes 
denoted $\Omega$).} 
\begin{equation}
\label{eq:intwf}
\langle \beta_l,\vec{r},s |\Psi_{\rm int} \rangle = \Phi(\beta_l) 
\psi_{K}(\vec{r},s) ~.
\end{equation}
Here ${\cal N}$ is a normalization constant, $\hat{R}_2$ rotates the
$D$-functions and the intrinsic wave function by 180 degrees around the
$y$-axis, and $\hat{P}$ changes the sign of $\vec{r}$ and of the odd-multipole
${\beta_l}$'s in the intrinsic wave function.  As long as we do not allow
nonaxial core vibrations to be excited, we can, for the purposes of this 
paper, write the Schiff operator ${S}_z$ as
\begin{equation}
S_z =   D^{1*}_{00} (\theta_i) \hat{S}_{{\rm 
int}}(\hat{\beta}_l,\vec{r},\vec{\sigma}) ~,
\end{equation}
where $\hat{S}_{\rm int}$ is the intrinsic-frame operator, with only the $K=0$
component relevant because others do not generate collective excitations in
the absence of nonaxial deformation or vibration.  [For $\hat{V}_{PT}$, the
transformation to the intrinsic frame is trivial because that operator is
invariant under rotation.]  We have placed hats on the $\beta$'s and (on
$S_{\rm int}$) because at the quantum level they are operators that act on the
wave functions $\Phi(\beta_l)$ in eq.\ (\ref{eq:intwf}).  All of this
formalism can be justified at least in part through projected mean-field
calculations, in which the state is a function of all $A$ nucleon coordinates;
we will argue shortly that such calculations are necessary to answer questions
that arise in the simpler description.

In the simple particle-rotor model, matrix elements of one-body operators like 
$S_z$ are straightforward to calculate.  The coordinates of the individual 
nucleons in the core are integrated over in the intrinsic frame to give an 
operator that is a function of the collective coordinates and those of the odd 
particle, and can be applied to wave functions of these same coordinates.  
When one does the integration for the Schiff operator, assuming a nuclear 
charge density proportional to the mass density (so that the intrinsic dipole 
moment is zero\footnote{The collective contribution to the dipole moment is 
obviously hard to calculate precisely if in this approximation it is 
identically zero.}), the result is
\begin{equation} 
\label{eq:Sop} 
\hat{S}_{\rm int} = ZeR_0^3\frac{3}{20\pi} \sum_{l=2} \frac{(l+1) 
\hat{\beta}_l \hat{\beta}_{l+1}} {\sqrt{(2l+1)(2l+3)}} + S_{\rm int}(\rm 
s.p.)~, 
\end{equation} 
where the last term is the contribution of the valence particle, which in this
case can be neglected compared to the collective piece.  Without the hats, the 
first term in eq.\ (\ref{eq:Sop}) is just the classical Schiff moment of a 
deformed drop.

Interesting things happen when a core is both quadrupole and octupole
deformed (i.e.\ when $\Phi(\beta_l)$ is peaked around nonzero values of
$\beta_2$ and $\beta_3$).  [For a comprehensive review of the subject, see
ref.\ \cite{Butler}.]  The reflection asymmetry implies a double-well
potential in the coordinate $\beta_3$, which in turn means that the wave
functions with good parity will be linear combinations of functions peaked
around some $\beta_3$ and its negative.  If the barrier between the two wells
is high enough, the result will be parity-doubling; low-lying states will
have partners nearby with the same angular momentum but opposite parity.  This
means, for one thing, that there will be an intermediate state in
(\ref{eq:SM}) that enters with a small energy denominator and therefore a
large amplitude.  In fact, it is often reasonable to ignore all other states 
in 
the sum (see the shell-model result for $^{19}$F in fig.~3), mainly because of 
the energy denominator, but also because if the deformation is strong enough
the matrix element of the Schiff operator to that state is likely to be 
large.  The reason for that is that the doublets can be
viewed as the projection onto positive and negative parity of the same
reflection-asymmetric intrinsic state $|\Psi_{\rm int}\rangle$.

The existence of more than one state with the same intrinsic structure is
exactly the same phenomenon as the rotational bands associated with ordinary
reflection-symmetric deformation.  The matrix element of an operator between
the two states of the same doublet is proportional to the diagonal
intrinsic-state matrix element of the intrinsic operator, just as it
is for states within a rotational band.  For the operator $\hat{S}_{\rm int}$,
this diagonal matrix element is large; it is given roughly by the expression
in eq.\ (\ref{eq:Sop}) with the operators $\hat{\beta}_l$ replaced by the
values around which the wave function $\Phi(\beta_l)$ is peaked (i.e.\ by the
classical Schiff moment of the deformed asymmetric core), with coherent
contributions from all the nucleons in it\footnote{As pointed out in refs.\
\cite{A2} and \cite{A3}, the collective enhancement of the dipole moment is
much smaller (zero in fact if the charge distribution is proportional to
the mass distribution) because the
dipole moment is measured from the center of mass.}.  Thus, from eq.\
(\ref{eq:SM}), the physical ground-state Schiff moment $S$ that determines the
atomic electric dipole moment should be approximately
\begin{equation}
S \approx -2 \frac{J}{J+1} 
S_{\rm int} \frac{\langle \widetilde{\Psi}_{0} | \hat{V}_{PT} | \Psi_0 
\rangle}{\Delta E}~,
\end{equation}
where $|\widetilde{\Psi}_0 \rangle$ is the opposite-parity partner of the 
ground
state $|\Psi_0\rangle$, $\Delta E$ is the energy difference between the two
states, and $S_{\rm int}$ is the  intrinsic Schiff moment (the $J$-dependent
factor is from the Euler-angle integration). The conclusion from all this is
that large intrinsic Schiff moments and small energy denominators should make
atoms with octupole-deformed nuclei especially sensitive tests of 
P,T-violation
in the nucleon-nucleon interaction.

Just how sensitive the tests will be depends on the matrix element of the
interaction $\hat{V}_{PT}$, to which we now turn.  We will supply more detail 
in addressing this subject because the treatment of ref.\ \cite{A3} is 
incomplete and not entirely accurate. We assume that the pion is responsible 
for transmitting most of the T-violating force from one nucleon to 
another \cite{town}.  The nucleon-nucleon interaction produced by one-pion 
exchange then has the general form\cite{Haxton,herczeg}
\begin{eqnarray}
\label{eq:pion}
\hat{V}_{PT}(12) & = & \left\{  \begin{array}{l}
                         (\vec{\sigma_1} - \vec{\sigma_2}) \cdot
                        (\vec{r_1}-\vec{r_2}) \left[ C_0 \vec{\tau_1} \cdot
                         \vec{\tau_2} + C_1 (\tau_{1z}+\tau_{2z}) + C_2        
                         (3\tau_{1z}\tau_{2z} - \vec{\tau_1} \cdot 
                         \vec{\tau_2} ) \right] \\
                         +  C_1 (\vec{\sigma_1}+\vec{\sigma_2}) 
                         \cdot (\vec{r_1}-\vec{r_2})
                         (\tau_{1z}-\tau_{2z}) \end{array}  \right\} 
                           \nonumber \\
&\times& \frac{{\rm exp}(-m_{\pi} |\vec{r_1}-\vec{r_2}|)}{m_{\pi}
|\vec{r_1}-\vec{r_2}|^2} \left[ 1+\frac{1}{m_{\pi}|\vec{r_1}-\vec{r_2}|}
\right]~,
\end{eqnarray} 
where the $C_i$'s label isoscalar, isovector, and isotensor contributions.  As
long as we excite no particles out of the core, P,T-violating interactions
between these particles sum to zero (see ref.\ \cite{A3} for discussion), and
we need worry only about the interactions between the valence particle and
those in the core.  In what follows, we take the total mass density to be 
proportional to the charge density, and in fact {\em from now on use the 
symbol $\rho$ to represent the mass density}.  This assumption means that the 
terms with different isospin structure in eq.\ (\ref{eq:pion}) enter in 
similar ways.  Taking the range of the pion to be very short (a decent
approximation\cite{vogel}), summing eq.\ (\ref{eq:pion}) over the particles in
the core, and assuming the neutron and proton densities to be equal gives the
effective nucleon-core interaction
\begin{equation}
\label{eq:sppot}
\hat{U}_{PT} = \eta \frac{G}{2 m \sqrt{2}} \vec{\sigma} \cdot \vec{\nabla} 
\hat{\rho} 
~,
\end{equation}
where (again) $\hat{\rho}$ is the core mass-density operator, $G$ is the Fermi
constant, and $m$ is the nucleon mass (these factors are inserted to follow
convention).  The dimensionless parameter $\eta$ then depends on the coupling
strengths $C_i$ of the two-body interactions and the isospin of the nucleus.
Despite its slight dependence on nuclear structure, this parameter is often
taken as a ``heuristic'' fundamental quantity.  The one-body approximation
given by eq.\ (\ref{eq:sppot}) is slightly different from the simpler and more
phenomenological one we used in our discussion of light nuclei.

The form of eq.\ (\ref{eq:sppot}) makes the Schiff moment unpleasantly
sensitive to the distribution of spin near the nuclear surface, where
$\vec{\nabla} \rho$ is largest.  The results of ref.\ \cite{A3} sometimes
differ by factors of several from those of ref.\ \cite{A2}, primarily because
of differences in the valence single-particle wave function, which carries all
the nuclear spin $\vec{s}$ in the particle-rotor model.  Only a significantly
more sophisticated calculation (which we advocate) will reduce this
uncertainty.  We therefore do not present our own complete
``particle-rotor-model-with-octupole-deformation'' calculations in this
section, but instead use that model in its simplest form, together with
qualitative arguments, to identify a few systematic effects overlooked in the
existing calculations.  Our estimate of the size of these effects is obviously
uncertain, but indicates what can be expected in more sophisticated
calculations.  The new physics always tends to lessen the enhancement.

To see the what was neglected in refs.\ \cite{A2,A3} we follow ref.\ 
\cite{Bohr} and expand the density in the deformation parameters:
\begin{eqnarray}
\label{eq:density}
\rho(\vec{r}) & \approx & \rho_0(r - R(\theta,\phi) + R_0) \nonumber \\
& \approx & \rho_0(r) - R_0 \rho_0'(r) \sum_l \beta_l Y_{l,0}
+ 1/2 R_0^2 \rho_0''(r) (\sum_l \beta_l Y_{l,0})^2 + \cdots~,
\end{eqnarray}
where now $\rho_0(r)$ is the bare (spherical) ground-state mass density of the 
core, a constant up to radius $R_0$ in the simplest version of the liquid-drop 
picture. The operator $\vec{\sigma} \cdot \vec{\nabla} \hat{\rho}$ in
$\hat{U}_{PT}$ therefore depends on the $\hat{\beta}$'s, and, contrary to the 
statements in ref.\ \cite{A3}, cannot be broken up into pseudoscalar pieces 
that act separately on the core and particle. In fact, from eq.\ 
(\ref{eq:density}) we have
\begin{eqnarray}
\label{eq:expansion}
\vec{\sigma} \cdot \vec{\nabla} \hat{\rho} &= &\vec{\sigma} \cdot \vec{\nabla} 
\rho_0
 \nonumber\\
&+& R_0 \sum_l \hat{\beta}_l \left[ \sqrt{\frac{l+1}{2l+1}} \left( 
\frac{d}{dr} - \frac{l}{r} \right)\rho_0'
[Y_{l+1} \sigma]^l_0 - \sqrt{\frac{l}{2l+1}}
 \left( \frac{d}{dr} + 
\frac{l+1}{r} \right) \rho_0' [Y_{l-1} \sigma]^l_0 \right]  \nonumber \\
&-& 1/2 R_0^2 \sum_{l,l',L} \hat{\beta}_l \hat{\beta}_{l'}
\sqrt{\frac{(2l+1)(2l'+1)}{4\pi (2L+1)^2}}\langle l0,l'0 | L0 \rangle 
\nonumber \\
& & \times \left[\sqrt{L+1} \left(\frac{d}{dr}-\frac{L}{r}\right) 
\rho_0''
[Y_{L+1}\sigma]^L_0 -\sqrt{L} \left( \frac{d}{dr}+\frac{L+1}{r} \right)
\rho_0'' [Y_{L-1}\sigma]^L_0 \right] ~,
\end{eqnarray}
where the small square brackets indicate angular-momentum coupling. Reference 
\cite{A3} considers only the first term in this expression, which changes the 
parity of the single particle and leaves the core alone.  The terms with odd 
powers of $\hat{\beta}_3$ can do the opposite, however.

Including these other terms is important because in the strong-coupling
limit of the particle-rotor model they tend to cancel the first term.  The
reason is hinted at in ref.\ \cite{A3}, where it is argued that Schiff moments
between close-lying states are suppressed when deformation is rigid and
symmetric.  Following ref.\ \cite{Sushkov}, the authors note that to the
extent that the density is proportional to the strong one-body potential
$\hat{U}_{\rm strong}$ felt by the odd particle and that the spin-orbit force
is negligible, matrix elements of $\vec{\sigma} \cdot \vec{\nabla} \hat{\rho}$
between two states should be proportional to the energy difference between
those states, and therefore very small for close-lying doublets.  The reason
is that under these circumstances
\begin{equation}
\vec{\sigma} \cdot \vec{\nabla} \hat{\rho}\, \propto \, \vec{\sigma} \cdot 
\vec{\nabla} \hat{U}_{\rm strong} = i [\vec{\sigma} \cdot \vec{p}, 
\hat{U}_{\rm strong}] = i [\vec{\sigma} \cdot \vec{p},\hat{H}(\rm s.p.)] ~,
\end{equation}
so that for two opposite-parity states labeled $a$ and $b$ with the same core 
structure, 
\begin{equation}
\langle \Psi^a_{JMK,p} |\hat{U}_{PT} |\Psi^b_{JMK,-p} \rangle \propto
\langle \psi_{a,K} | 
\vec{\sigma} \cdot \vec{\nabla} {\rho} | \psi_{b,K} \rangle
\propto \epsilon_a - \epsilon_b~,
\end{equation}
where the $\epsilon$'s are single-particle energies.  The authors then 
argue that complications associated with asymmetric deformation eliminate this 
effect, but in the strong-coupling limit the situation is even worse because 
now the two states 
$|\Psi^a_{JMK,p} \rangle$ and $|\Psi^b_{JMK,-p} \rangle$ have the 
same intrinsic 
structure.  As mentioned above, the matrix element of any operator between two 
such states is proportional in the strong-coupling limit to the 
intrinsic-state expectation value of the operator.  In nuclei with strong 
octupole deformation, we therefore have
\begin{equation}
\label{eq:spexp}
\langle \Psi_0 | \hat{V}_{PT} | \widetilde{\Psi}_0 \rangle \longrightarrow 
\langle \Psi_{\rm int} | \hat{U}_{PT} | \Psi_{\rm int} \rangle = \eta \frac{G}
{2 m \sqrt{2}} \langle \psi_{K} | \vec{\sigma} \cdot \vec{\nabla} 
{\rho} | \psi_{K} \rangle~,
\end{equation}
which, according to the argument above, should vanish.  The estimates of 
refs.\ \cite{A2,A3} apparently neglect the terms in $\rho$ containing the 
$\beta$'s that make the shapes of the density distribution and the potential 
similar, and so do not take this effect into account.  

Of course the spin-orbit force is not negligible and the intrinsic density is
not exactly proportional to the mean field, so the cancellation will not be
complete.  To get a handle on how much the terms containing the $\beta$'s
affect the matrix element of $\hat{U}_{PT}$, we consider the ratio of the
matrix element in eq.\ (\ref{eq:spexp}) with the terms included to that
without them (the latter being the a simplified version of the quantity
calculated in refs.\ \cite{A2,A3}) for a large number of single-particle
orbits.  We use a deformed harmonic oscillator as a potential,
\begin{equation} \label{eq:defho} V(\vec{r}) = -m \omega r^2 \sum_l \beta_l
Y_{l,0}~, \end{equation} with deformations $\beta_2$, $\beta_3$, and $\beta_4$
equal to those from Table I of ref.\ \cite{A3} (we ignore higher multipoles
and neglect pairing).  We take the density to be constant inside the liquid
drop and zero outside, a distribution that has the same angular shape as the
potential, but a significantly different radial dependence.  Figure 4 shows
the absolute value of the ratio of $\langle \psi_{K} | \vec{\sigma} \cdot
\vec{\nabla} \hat{\rho} | \psi_{K} \rangle$ to the same matrix element without
the $\beta$-dependent terms, for all $K=1/2$ and $K=3/2$ single-particle
levels in $^{225}$Ra below 8 $\hbar\omega$ of single-particle energy.  The new
terms generally have the opposite sign from that of the $\beta$-independent
term, and the sum is usually less than the first term by itself.  The average
cancellation is less for $j=1/2$ states than for states with larger spin.
Whether the decrease will be stronger or weaker in realistic calculations is
an open question.  In self-consistent mean-field calculations, though, there
is obviously a correlation between the density and the spin-independent part
of the field, and we could well see a significant effect\footnote{We should 
add
that the decoupling of the particle and core\cite{Leander1}, taken into
account neither here nor in ref.\ \cite{A3}, could act like the spin-orbit
interaction to mitigate the suppression.}.

\begin{figure}[htb]
\begin{center}
\includegraphics[width=14cm,height=12cm]{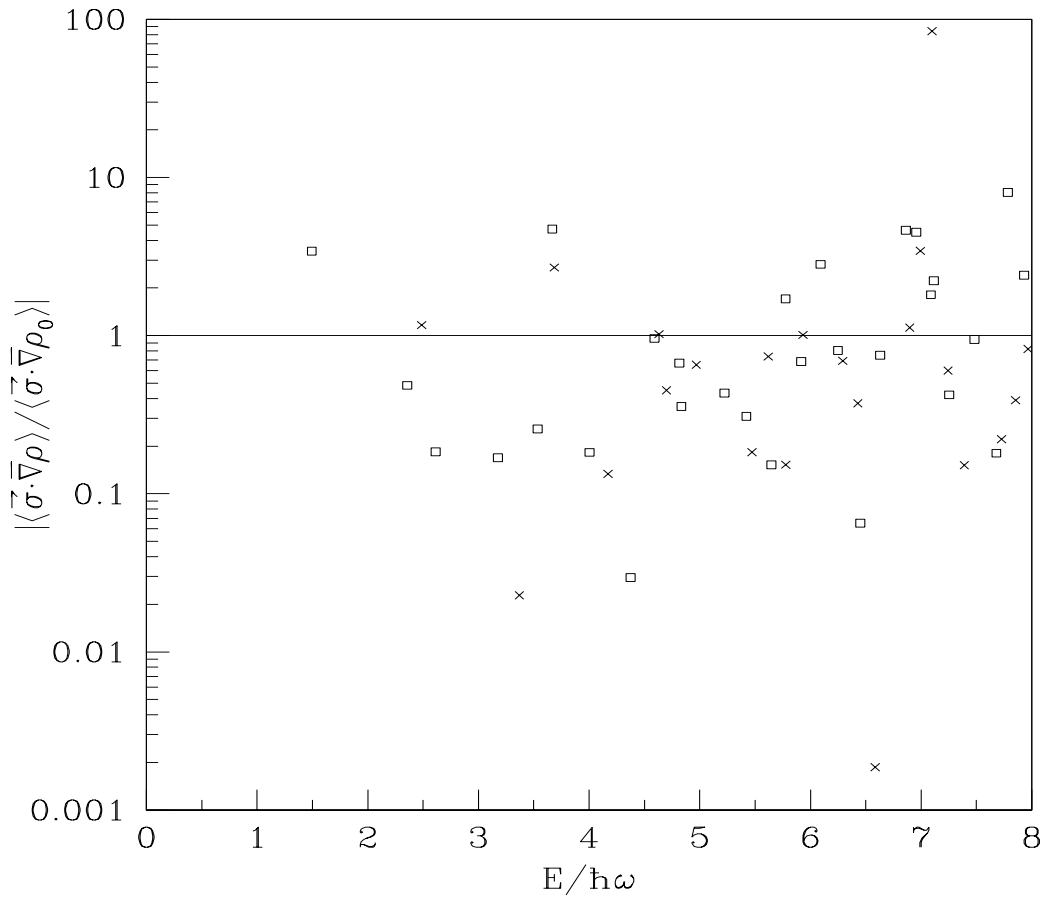}
\end{center}
\caption[Figure 4]{The ratio $|\vec{\sigma} \cdot \vec{\nabla} \rho|$ to 
$|\vec{\sigma} \cdot \vec{\nabla} \rho_0|$ (see text) for all $K=1/2$ (boxes) 
and 3/2 (crosses) states in a deformed asymmetric potential for $^{225}$Ra.  
The ratio is usually less than one.}
\label{f:4}
\end{figure}

Other physics neglected in the earlier work will also have an effect.  The
first term in eq.~(\ref{eq:expansion}) ($\vec{\sigma}\cdot \vec{\nabla} \rho_0
= (1/r)\rho_0' \vec{\sigma} \cdot \vec{r}$) is the spin-flip analog of the
electric dipole operator (with a difference only in radial form and isospin).  
We refer to
fig.~2, where the strength of the related operator $\vec{\sigma} \cdot
\vec{r}$ in the isovector and isoscalar channels is plotted for $^{19}$F.  As
noted in section III, the strength is clearly concentrated in resonances at
about the same energy as the giant dipole resonance.  A simple
argument\cite{Leander2} with a schematic residual interaction in RPA shows
that the existence of the giant dipole resonance of the core suppresses E1
transitions between low-lying single-particle states by a factor of 3 or 4
that depends only on the energy of the resonance and the energy at which
strength would be centered if there were no resonance.  The low-lying
transitions induced by $\vec{\sigma}\cdot \vec{\nabla} \rho_0$ should be
suppressed by roughly the same amount because the operator
$\vec{\sigma}\cdot\vec{r}$ is so much like $\vec{r}$ (again, see fig.~2).  In
many-body perturbation theory, this effect can be understood as core
polarization:  the residual strong interaction can create a collective $0^-$
p-h pair, at the same time changing the parity of the valence particle.  The
contribution to the Schiff moment coming from the annihilation of the pair by
$\hat{U}_{PT}$ nearly cancels the contribution coming from the direct action
of $\hat{U}_{PT}$ on the valence particle itself.  In any event, the
resonances were completely neglected in refs.\cite{A2,A3}, and the
single-particle matrix elements in those papers should therefore probably be
three or four times smaller.

The very large Schiff moments may be saved, however, by the {\em combination}
of the two new effects, even though each reduces Schiff moments when added to
the calculations of refs.\ \cite{A2,A3} in isolation.  The spin-flip giant
resonance should not affect those parts of $\hat{U}_{PT}$ that contain
$\beta_3$ and operators like $[Y_2 \sigma]^3_0$ that do not change the parity
of the valence particle.  The suppression by the residual interaction of one
part of the Schiff moment by a factor of three without any effect on another
part may upset any balance between the two at the mean-field level produced by
the similarity between the density and the potential.  We will need accurate
microscopic calculations to test the existence of both effects and the extent
to which they offset one another.

Such calculations are within the range of today's mean-field technology, and
in even-even nuclei they have already been carried out\cite{egido1,martin},
confirming the large intrinsic Schiff moments in the radium
isotopes\cite{egido2}.  In self-consistent (e.g.\ Skyrme-HFB) calculations the
effects of the residual interaction on the ground state are minimized.
Therefore, not only is the relationship between the mean field and the density
likely to be most accurate in this case, but it is also least likely to be
vitiated by corrections to the single-particle picture.  In odd nuclei, the
first-order core polarization, which has to be treated as a correction to
particle-rotor/Nilsson models, is built into the mean-field; 1-particle-1-hole
excitations of the core do not mix with the ground state.  This fact, together
with a realistic two-body interaction that contains all multipole-multipole
terms, has implications for corrections to low-lying transitions from
resonances, as well as for the density.  The incorporation of core
polarization means that the interplay between collective excitations and
low-lying states is already apparent at the mean-field level, or in other
words that the usual first-order particle-phonon mixing that reduces low-lying
single-particle transitions need not be treated by other means (e.g.\ the
RPA).  We should therefore be able at the mean-field level to go a long way
towards quantifying the influence of shape and resonances on Schiff moments in
octupole-deformed nuclei.

\section{Collective Schiff moments from octupole vibrations}

\indent

The question of whether the light actinides are octupole deformed has a long
history.  In fact the question is not entirely physical --- it's really about
the economy of one collective-model basis versus another --- and it should not
matter so much whether the low-lying states in a nucleus are best described as
the rotation of an octupole-deformed shape or as a strong low-lying octupole
vibration around a rotating quadrupole shape.  
Collective Schiff moments arise in either scheme.  This fact should not be 
surprising in light of our calculations in $^{19}$F.  To see how it falls out 
of the collective picture, we assume that the nuclear core has no 
equilibrium octupole deformation (i.e.\ $\langle \beta_3 \rangle = 0$) and 
write the operator $\hat{\beta}_3$ in terms of creation and annihilation 
operators:
\begin{equation}
\label{eq:phonon}
\hat{\beta}_3 \propto b^{\dag} + b~,
\end{equation}
where $b^{\dag}$ creates an octupole phonon with (intrinsic) magnetic quantum 
number $K = 0$.  It is then clear from eq.\ (\ref{eq:Sop}) that the Schiff 
operator acting on a quadrupole-deformed state with no octupole phonons will 
create an excited state with one phonon. The terms in $\hat{U}_{PT}$ (see eq.\ 
(\ref{eq:expansion})) that are proportional to $\hat{\beta}_3$ can then 
destroy the phonon, reconnecting the one-phonon state to the ground state and 
generating a collective Schiff moment through eq.\ (\ref{eq:SM}).  

To see how big such a moment would be we need to know the matrix element of
$\hat{\beta}_3$ between states with zero and one phonons.  As can be seen from
eq.\ (\ref{eq:phonon}), this quantity is just the zero-point root-mean-square 
deformation, ($\sqrt{\langle \hat{\beta}_3^2 \rangle}$), which we will call
$\bar{\beta}_3$. [In other words, $\bar{\beta}_3$ measures the spread in 
$\beta_3$ of the intrinsic core wave function $\Phi(\beta_l)$.]  This quantity 
can be estimated from the collective (vibrational) B(E3) transition 
in an even-even neighbor.  Using eq.\ (\ref{eq:density}) to lowest order in 
$\bar{\beta}_3$ one finds\cite{Spear}
\begin{equation}
\label{eq:E3}
B(E3)_{0^+ \rightarrow 3^-} = (3/4\pi)^2 (ZeR_0^3)^2 \bar{\beta}_3^2~.
\end{equation}
The important point is that if a collective vibration is soft the r.m.s.\
deformation $\bar{\beta}_3$ can be as large as the value around which the
wave-function is peaked in octupole-deformed nuclei, and the intrinsic Schiff
moment can therefore be just as large as well.  In the laboratory (physical)
Schiff moment, there is an additional factor of $\bar{\beta}_3$ coming from
the annihilation of the phonon by $\hat{U}_{PT}$, so that naively we expect
the moment to depend on the deformation parameters in the combination $\beta_2
\bar{\beta}_3^2$, where an unbarred $\beta$ is the value around which the
deformed wave function is peaked.  The relevant quantity for octupole-deformed
nuclei is $\beta_2 \beta_3^2$ (see ref.\ \cite{A3} for a discussion of why),
so that if the r.m.s.\ octupole deformation $\bar{\beta}_3$ in a vibrational
nucleus is comparable to the static value $\beta_3$ of the deformation in an
octupole-deformed nucleus, any differences in Schiff moments come from the
energy denominator, single-particle structure, or other core excitations, not
from the difference between deformation and vibration.  We will refine this
statement shortly.

First, however, we note that the terms in $\hat{U}_{PT}$ that don't contain
$\hat{\beta}_3$ are usually even more important than those just discussed,
even though they don't alter the number of phonons, because the zero- and
one-phonon states mix through the residual strong particle-core interaction.
The approximate form of this coupling can be derived in many ways; one is to
examine the change in energy under a small deformation of the core.  Not
surprisingly, for an oscillator single-particle potential this leads to the
same interaction that appears in the octupole-deformed potential of the 
strong-coupling scheme (see eq.\ (\ref{eq:defho}):
\begin{equation}
\hat{V}_{\rm coupl}  = - m \omega^2 \hat{\beta}_3 r^2 Y_{3,0} ~,
\end{equation} 
where $\omega$ is the oscillator energy of the (symmetric) potential, 
$\hat{\beta_3}$ acts on the core, and $r^2 Y_{3,0}$ acts on the particle.  
Denoting a state with $n$ phonons and a particle in orbit $\psi_{a,K}^{p}$ 
by $| n,\psi_{a,K}^{p}\rangle$, we have for the matrix element of the 
interaction 
$\hat{V}_{\rm coupl}$ between excited states with one phonon and the 
unperturbed ground state (assuming just for illustration that the ground state 
has positive parity):
\begin{equation}
\langle 1, \psi_{b,K}^{-} | V_{\rm coupl} | 0, \psi_{a,K}^{+} \rangle  = -m 
\omega^2 
\bar{\beta}_3 \langle \psi_{b,K}^{-} | r^2 Y_{3,0} | \psi_{a,K}^{+} \rangle ~.
\end{equation}

With a value for $\bar{\beta}_3$ from an appropriate $B(E3)$, we can use the
``intermediate''-coupling scheme of ref.\ \cite{Leander1} to diagonalize 
$\hat{H}
\equiv \hat{H}({\rm s.p.}) + \hat{H}({\rm phonon}) + \hat{V}_{\rm coupl}$ 
separately in positive- and negative-parity bases ($\hat{H}({\rm phonon})$ 
just contains the diagonal vibrational energies of the zero- and one-phonon 
states), so that the ground state has the form
\begin{equation}
\label{eq:gs}
|\Psi_0 \rangle = \sum_i A_i | 0, \psi_{i,K}^{+} \rangle + \sum_j B_j | 1, 
\psi_{j,K}^{-} \rangle~,
\end{equation}
and the excited states of opposite parity have the form 
\begin{equation}
\label{eq:excited}
| \Psi_l \rangle = \sum_i C_{l,i} |1,\psi_{i,K}^{+} \rangle + \sum_j D_{l,j} 
| 0, \psi_{j,K}^{-} \rangle~,
\end{equation}
where the $\psi_{i,K}^{+}$ and $\psi_{j,K}^{-}$ label single-particle states 
around
the Fermi surface, and we are still ignoring nonaxial vibrations.  The terms 
in $\hat{U}_{PT}$ that are independent of
$\hat{\beta}_3$ connect the first terms in eq.\ (\ref{eq:gs}) to the second in
eq.\ (\ref{eq:excited}) and vice versa.
The Schiff operator affects the core,
connecting the first term in eq.\ (\ref{eq:gs}) to the first in eq.\
(\ref{eq:excited}), and the second to the second, effectively replacing
$\hat{\beta}_3$ in eq.\ (\ref{eq:Sop}) by $\bar{\beta}_3$.  In this way the
spherical $\hat{\beta}_3$-independent part of $\hat{U}_{PT}$ (the only part
considered in refs.\ \cite{A2,A3}) can also generate a collective Schiff
moment.

It is possible to use the intermediate-coupling scheme even as the phonon
energy goes to zero and octupole deformation sets in.  In that case, because
the single-particle Hamiltonians in the two schemes are the same, energies and
matrix elements should not depend strongly on which scheme is
used\footnote{They will not be identical because in the intermediate-coupling
scheme some of the states are particles and some holes, and single-particle
excitation energies are measured with respect to the Fermi
surface\cite{Leander1}.}.  One implication (which is a stronger version of a
remark made above) is that if the dynamic $\bar{\beta}_3$ associated with the
vibration is comparable to the static $\beta_3$ in an octupole-deformed
nucleus, and if the energy of the octupole phonon is small compared to typical
single-particle splittings or nonaxial core-excitation energies, the {\em
only} major difference between Schiff moments in the two cases is the energy
denominator in eq.\ (\ref{eq:SM}).  To see this, one can imagine treating the
phonon as a ``decoupling'' perturbation (along with the Coriolis interaction)
in the strong-coupling scheme, as is done in ref.\ \cite{Leander1}.  Although
the diagonal matrix elements of the perturbation cause energy shifts, wave
functions are only affected by the off-diagonal matrix elements with bands
built on higher single-particle states or other kinds of vibrations.  Thus
wave functions, transition amplitudes, etc., will not undergo large changes
until the energy of the phonon approaches those of other excitations.  For the
intermediate-coupling scheme, this means that the matrix elements of $S$ and
$\hat{V}_{PT}$ connecting ground states to low-energy octupole phonons should
not undergo radical change once the phonon is low enough in energy, and that
nothing special will happen in the limit that the phonons have zero energy and
the core develops static deformation.  Of course if the phonon lies high in
the spectrum, the matrix elements can be very different from the static limit,
and one must carry out the intermediate-coupling calculation to get a handle
on the size of the Schiff moment induced by vibrations.

We have done just that in several quadrupole-deformed nuclei, taking
vibrational $\bar{\beta}_3$'s and phonon energies from tabulations of nearby
even-even nuclei\cite{Spear}, and again neglecting pairing.  In $^{199}$Hg,
the most accurately measured isotope at present, we use $\bar{\beta}_3$ = .09,
a phonon energy of 3 MeV (both taken from an E3 transition in $^{204}$Hg,
which may have a larger $\bar{\beta}_3$ than $^{199}$Hg), and quadrupole and
hexadecupole parameters from ref.\ \cite{Moller} (for this simple estimate we
ignore the fact that this nucleus is probably very soft).  The resulting
Schiff moment is $8 \times 10^{-9} \eta$ e ${\rm fm}^3$, about half of the
estimate from ref.\ \cite{Flambaum} that includes no nuclear correlations of
any kind.  In a nucleus like this, moreover, with a relatively high-energy
phonon, the nonaxial octupole vibrations will lie nearby in energy and can be
expected to contribute comparable amounts.  When all is said and done,
vibrations may turn out to be the dominant contribution to the Schiff moment
in $^{199}$Hg, and they clearly should be included in any realistic
calculation.  Such a calculation has never been done, but is crucial if we
want a reliable assessment of the advantages offered by nuclei with strong
octupole correlations.  Here we need a good microscopic treatment of all kinds
of vibrations, including the very soft $\gamma$ quadrupole mode, and must
obviously go beyond mean-field theory.  A shell-model calculation may be
possible.

An example of a large vibrational Schiff moment is in the nucleus $^{239}$Pu.
This isotope has several features that make it attractive for experiment, in
particular a spin-1/2 ground state (to eliminate quadrupole effects in a
magnetic field) and a long half-life compared to the light actinides [The
drawback is in its electronic structure, which is more complicated than that
of Radium].  The collective E3 in $^{238}$Pu gives $\beta_3=.09$, and the
nucleus has large quadrupole and hexadecupole deformations ($\beta_2 = .223$,
$\beta_4 = .095$).  Together with the high value of $Z$, this makes the
intrinsic Schiff moment very large.  The phonon lies at 470 keV, about 8 times
higher than the lowest state in $^{225}$Ra, but the large intrinsic Schiff
moment compensates in part.  Our calculations, with only the
$\hat{\beta}$-independent terms included in $\hat{\rho}$ (as in ref.\
\cite{A3}) give a laboratory Schiff moment of $7 \times 10^{-7} \eta ~e~ {\rm
fm}^3$, a value a few times smaller than the results of ref.\ \cite{A3} for
most of the light actinides.  When we include the $\hat{\beta}$-dependent
terms, this number goes up to $4 \times 10^{-6} \eta ~e~ {\rm fm}^3$, which is
300 times the estimate for Hg in ref.\ \cite{Flambaum} and comparable to
the results of ref.\ \cite{A3} for $^{225}$Ra.  These calculations are far
from perfect; we had to push the energy of the octupole phonon well above the
value from $^{238}$Pu to get the energy of the first excited state in
$^{239}$Pu right.  The uncertainty in the results is therefore quite large and
we need microscopic calculations here too.  But the intrinsic Schiff
moments will of nuclei with low-lying octupole vibrations will clearly be
collective, and some may of these nuclides may be easier to investigate
experimentally than the short-lived Radium isotopes.

\section{Conclusion}

\indent

The size of Schiff moments in nuclei with octupole correlations is determined 
by three factors:  intrinsic Schiff moments, energy denominators, and the 
matrix elements of $\hat{V}_{PT}$.  In their discussion of the first two of 
these, refs.\ \cite{A2,A3} are on rather firm ground; it is hard to imagine, 
for example, that the matrix elements between parity doublets of the Schiff 
operator are radically different from those estimates, and as we have pointed 
out, even nuclei without asymmetrically deformed cores can benefit from the 
same mechanism.  The third factor is far trickier, however.  

The particle-core calculations reported both here and in earlier work can only
supply a gross estimate of the matrix element of $\hat{V}_{PT}$.  The mixing
that that interaction induces depends sensitively on the valence
single-particle wave function at the nuclear surface, where $\vec{\nabla}
\rho$ is largest.  Truly microscopic calculations will give better valence
wave functions and, if they are self-consistent, will also better represent
the correlation between density and mean field, and incorporate the effects of
resonances caused by the residual interaction.  In vibrational nuclei it will
be necessary to go a little further, but even there mean-field calculations 
will shed light on the issues we've discussed.

Finally, for experimentalists to draw strong conclusions about enhancements
over ${}^{199}$Hg, better calculations in that nucleus must be done as well.
It is conceivable that the atomic dipole moment of $^{225}$Ra is 400 or more
times larger than that of $^{199}$Hg (this is the figure reported in ref.\
\cite{A3}), but we have pointed to physical effects that could make the Schiff
moment in ${}^{199}$Hg a few times larger than earlier calculations indicate
and the Schiff moments in the light actinides somewhat smaller than suggested
by the calculations of refs.\ \cite{A2,A3}, even within the same model.  The
machinery of modern nuclear structure theory, which is powerful enough to
provide reasonably accurate estimates of the moments in both kinds of nuclei,
should be used as soon as possible to provide experimentalists firm
predictions for the enhancement they can expect in difficult experiments with
radioactive nuclei.

\section{Acknowledgements}

\indent

We thank J.L.\ Egido for useful calculations of Schiff moments in the
even-even light actinides.  We are grateful to W.\ Nazarewicz, I.B.\
Khriplovich, P.\ Herczeg, and S.\ Lamoreaux for helpful comments about the
manuscript.  The work of J.E.\ was supported by the U.S.\ Department of Energy
under grant DE--FG02--97ER41019, and the work of J.L.F.\ and A.C.H.\ was
performed under the auspices of the United States Department of Energy.

\end{document}